\documentclass[twoside]{ilcws08}
\usepackage[latin1]{inputenc}
\usepackage[dvips]{graphicx,epsfig,color}
\usepackage{wrapfig,rotating}
\usepackage{amssymb,amsmath,array}

\begin{document}
\title{Calculating gluon one-loop amplitudes numerically}
\author{Jan-Christopher Winter and Walter T.\ Giele
\thanks{J.W.\ thanks the Aspen Center of Physics, where parts of this
  work were accomplished during the Summer Workshop ``LHC: BSM Signals
  in a QCD Environment''.}
\vspace{.3cm}\\
Fermi National Accelerator Laboratory\\
P.O.\ Box 500, Batavia, IL 60510 - U.S.A.\\
}
\maketitle

\begin{flushright}
  \vspace*{-44mm}
  {\small FERMILAB-PUB-09-029-T}\\[36mm]
\end{flushright}

\begin{abstract}
This note reports on an independent implementation of calculating
one-loop amplitudes semi-numerically using generalized unitarity
techniques. The algorithm implemented in form of a C++ code closely
follows the method by Ellis, Giele, Kunszt and
Melnikov~\cite{Ellis:2007br,Giele:2008ve}. For the case of gluons, the
algorithm is briefly reviewed. Double-precision results are presented
documenting the accuracy and efficiency of this computation~\cite{url}.
\end{abstract}

\section{Introduction}
\label{sec:intro}
An automated next-to-leading order generator for Standard Model
processes is highly desirable \cite{Bern:2008ef}. With recent
developments of generalized unitarity \cite{Bern:1994zx} and
parametric integration methods \cite{Ossola:2006us} such a generator
seems to be within reach
\cite{Ellis:2007br,Giele:2008ve,Giele:2008bc,Ossola:2007ax}.
A first crucial step is the development of stable and fast algorithms
for evaluating one-loop amplitudes through generalized unitarity cuts.
A C++ code is especially of interest because of the ease with which it
can be integrated in leading-order generators such as
C\scalebox{0.8}{OMIX} \cite{Gleisberg:2008fv}. The leading-order code
will then be used to compute the cut graphs. Eventually, such a
retrofitted generator will be able to generate all necessary
amplitudes for a next-to-leading order Monte Carlo program for any
Standard Model process of interest to the collider experiments.

\section{Construct one-loop by tree-level amplitudes -- algorithm
  in brief}
\label{sec:algo}
The full $N$-gluon one-loop amplitude can be constructed from the
leading colour-ordered amplitudes \cite{Bern:1990ux}, which can be
calculated by $A^{[1]}_N(\{p_i,\kappa_i\})=A^{\rm cc}_N+R_N$ depending
on external momenta $p_i$ and polarizations $\kappa_i$. The
cut-constructible part reads
\begin{equation}
  A^{\rm cc}_N\;=
  \sum_{[i_1|i_4]}d^{(0)}_{i_1i_2i_3i_4}I^{(4-2\epsilon)}_{i_1i_2i_3i_4}+
  \sum_{[i_1|i_3]}c^{(0)}_{i_1i_2i_3}I^{(4-2\epsilon)}_{i_1i_2i_3}+
  \sum_{[i_1|i_2]}b^{(0)}_{i_1i_2}I^{(4-2\epsilon)}_{i_1i_2}
\end{equation}
employing the short-hand notation $[i_1,i_M]=1\le i_1<i_2<\ldots<i_M\le N$,
and $M$\/ denotes the number of cuts. The master integrals are defined
as $I^{(D)}_{i_1\cdots i_M}=
\int d^D\ell\;(i\pi^{D/2}\,d_{i_1}\cdots d_{i_M})^{-1}$ and the
inverse propagators $d_i$ are functions of the loop momentum:
$d_i(\ell)=(\ell+q_i-q_{i_M})^2$ with $q_k=\sum^k_{j=1} p_j$. The
rational part $R_N$ is represented by
\begin{equation}
  R_N\;=\;
  -\sum_{[i_1|i_4]}\frac{d^{(4)}_{i_1i_2i_3i_4}}{6}+
  \sum_{[i_1|i_3]}\frac{c^{(7)}_{i_1i_2i_3}}{2}-
  \sum_{[i_1|i_2]}\left(\frac{(q_{i_1}-q_{i_2})^2}{6}\right)b^{(9)}_{i_1i_2}\,,
\end{equation}
cf.\ \cite{Giele:2008ve,Giele:2008bc}.
For each possible cut configuration $i_1\cdots i_M$, the box
($d^{(n)}_{i_1i_2i_3i_4}$), triangle ($c^{(n)}_{i_1i_2i_3}$) and
bubble ($b^{(n)}_{i_1i_2}$) coefficients appearing above are found as
solutions to the parametric form of the unintegrated ordered one-loop
amplitude,
\begin{equation}
  {\cal A}^{(D_s)}_N(\ell)\;=\;
  \sum_{[i_1|i_5]}\frac{\bar e^{(D_s)}_{i_1i_2i_3i_4i_5}(\ell)}
      {d_{i_1}d_{i_2}d_{i_3}d_{i_4}d_{i_5}}+
  \sum_{[i_1|i_4]}\frac{\bar d^{(D_s)}_{i_1i_2i_3i_4}(\ell)}
      {d_{i_1}d_{i_2}d_{i_3}d_{i_4}}+
  \sum_{[i_1|i_3]}\frac{\bar c^{(D_s)}_{i_1i_2i_3}(\ell)}
      {d_{i_1}d_{i_2}d_{i_3}}+
  \sum_{[i_1|i_2]}\frac{\bar b^{(D_s)}_{i_1i_2}(\ell)}{d_{i_1}d_{i_2}}
  \,.
\end{equation}
This decomposition of the integrand ${\cal A}^{(D_s)}_N(\ell)=
{\cal N}^{(D_s)}(\ell)/(d_1d_2\cdots d_N)$
has been generalized to higher (integer) dimensionality of the
internal particles, i.e.\ spin-polarization states and loop momenta
respectively have dimension $D_s$ and $D\le D_s$ (as required by
dimensional regularization). The $D_s$ dependence of the integrand can
be eliminated by taking into account that the numerator only linearly
depends on the spin-space dimension:
${\cal N}^{(D_s)}(\ell)={\cal N}_0(\ell)+(D_s-4)\,{\cal N}_1(\ell)$.
Moreover, only up to $5$-point terms (i.e.\ $M\le5$) need to be
included in the parametrization, since the loop momentum effectively
has $4+1$ components only:
${\cal A}^{(D_s)}_N(\ell)={\cal A}^{(D_s)}_N(\ell_1,\ldots,\ell_4,
[-\sum^D_{i=5}\ell^2]^{1/2})$. These two additions essentially are
sufficient to disentangle the rational part in the same way as
the cut-constructible part \cite{Giele:2008ve}.

The numerator functions of the $M$-point terms are polynomials
encoding the loop-momentum dependence residing in the space orthogonal
to the physical space defined by the external momenta. The orthogonal
space is spanned by the basis vectors $n_i$. The form of the
polynomials is richer compared to the four-dimensional case. They now
include the coefficients that determine the rational part, e.g.\ for
the box numerator, one finds
\begin{equation}
  \bar d_{i_1\cdots i_4}(\ell)\;=\;
  d^{(0)}_{i_1\cdots i_4}+\alpha_4\,d^{(1)}_{i_1\cdots i_4}+
  s^2_e\,[d^{(2)}_{i_1\cdots i_4}+\alpha_4\,d^{(3)}_{i_1\cdots i_4}]+
  s^4_e\,d^{(4)}_{i_1\cdots i_4}\,,
  \label{eq:dcff}
\end{equation}
where $\alpha_i=\ell\cdot n_i$ and $s^2_e=-\alpha^2_5-\ldots-\alpha^2_D$.
To solve for the coefficients of the numerator functions, loop momenta
$\ell=\ell_{i_1\cdots i_M}$ have to be constructed such that
$d_j(\ell_{i_1\cdots i_M})=0$ for $j=i_1,\ldots,i_M$. Then
\begin{eqnarray}
  \label{eq:ebar}
  \bar e^{(D_s)}_{i_1\cdots i_5}(\ell)&=&
  \mbox{Res}_{i_1\cdots i_5}\left[{\cal A}^{(D_s)}_N(\ell)\right]
  \;\equiv\;
  d_{i_1}(\ell)\cdots d_{i_5}(\ell)\ {\cal A}^{(D_s)}_N(\ell)
  \Bigr|_{d_{i_1}(\ell)=\cdots=d_{i_5}(\ell)=0}
  \,,\\
  \label{eq:dbar}
  \bar d^{(D_s)}_{i_1\cdots i_4}(\ell)&=&
  \mbox{Res}_{i_1\cdots i_4}\left[{\cal A}^{(D_s)}_N(\ell)\;-
    \sum_{[j_1|j_5]}\frac{\bar e^{(D_s)}_{j_1\cdots j_5}(\ell)}
	{d_{j_1}\cdots d_{j_5}}\right]\,,\quad\ldots\,.
\end{eqnarray}
Using the $n_i$ vectors, loop momenta fulfilling the constraints can
be generated: $\ell_{i_1\cdots i_M}=V_{i_1\cdots i_M}+
\sum^D_{i=M}\alpha_in_i$ where $\alpha^2_M=-V^2_{i_1\cdots i_M}-
\sum^D_{i=M+1}\alpha^2_i$.
In general, the solutions are found as complex momenta. The vectors
$V_{i_1\cdots i_M}$ reside in the physical space; they are constructed
from sums of external momenta as specified by the cuts, cf.\
\cite{Ellis:2007br}. By successively applying quintuple, quadruple,
triple and double $D_s$-dimensional unitarity cuts, all coefficients
can be determined. The cuts yield $M$\/ onshell propagators
factorizing the unintegrated one-loop amplitude into $M$\/ tree-level
amplitudes. Hence, the residues can be evaluated:
\begin{equation}
  \label{eq:residues}
  \mbox{Res}_{i_1\cdots i_M}\left[{\cal A}^{(D_s)}_N(\ell)\right]\;=
  \sum^{D_s-2}_{\{\lambda_1,\ldots,\lambda_M\}=1}
  \left\{
  \prod^M_{k=1}{\cal M}^{(0)}\left(\ell^{(\lambda_k)}_{i_k};
  p_{i_k+1},\ldots,p_{i_{k+1}};-\ell^{(\lambda_{k+1})}_{i_{k+1}}\right)
  \right\}
\end{equation}
where $\ell_{i_k}=\ell+q_{i_k}-q_{i_M}$ and the sum is over internal
polarization states.

\section{C++ implementation and results}
\label{sec:results}
The algorithm described above has been implemented in a new C++ code.
The only interface is to link the QCDLoop package \cite{Ellis:2007qk}
for evaluating the master integrals. The program is capable of
calculating the $A^{[1]}_N(\{p_i,\kappa_i\})$ amplitudes in double
precision in the four-dimensional helicity scheme. This allows for
crosschecks with the results obtained in
Refs.~\cite{Giele:2008bc,Lazopoulos:2008ex}.

The construction of the orthonormal sets of the $D-M+1$ basis vectors
and the $D_s-2$ polarization vectors follows the method outlined in
\cite{Giele:2008bc}. In addition, the $n_i$ vector generation
($i=M,\ldots,D$) has been set up such that basis vectors obtained for
large-$M$\/ cuts can be re-used for suitable lower-$M$\/ cuts. The
basic strategy, which has been implemented to find the coefficients,
is as follows: by using the freedom in choosing loop momenta, $m$\/
solutions and, therefore, algebraic equations, such as
eq.~(\ref{eq:dcff}), can be generated to solve for $n\le m$\/
coefficients.\footnote{In principle, an infinite number of equations
  can be generated to fit a fixed number of unknowns.}
First the dependence on $D_s$ is eliminated by computing:
$\mbox{Res}_{i_1\cdots i_M}[{\cal A}_N(\ell)]=
(D_s-3)\,\mbox{Res}_{i_1\cdots i_M}[{\cal A}^{(D_s)}_N(\ell)]-
(D_s-4)\,\mbox{Res}_{i_1\cdots i_M}[{\cal A}^{(D_s+1)}_N(\ell)]$.
Then higher-point terms are subtracted yielding numerator factors
$\bar e_{i_1\cdots i_5}(\ell)$ etc.\ that are independent of $D_s$,
i.e.\ eqs.~(\ref{eq:ebar}) and (\ref{eq:dbar}) work without the
$(D_s)$ label. For the coefficients of the cut-constructible part, one
can dispense with the determination of the $D_s$ dependence of the
residues and set $D=D_s=4$. This, in addition, leads to smaller
subsystems of equations, which can be solved separately, e.g.\
eq.~(\ref{eq:dcff}) simplifies to $\bar d_{i_1\cdots i_4}(\ell)\;=\;
d^{(0)}_{i_1\cdots i_4}+\alpha_4\,d^{(1)}_{i_1\cdots i_4}$.
The tree-level amplitudes needed to obtain the residues in
eq.~(\ref{eq:residues}) are calculated with Berends--Giele recursion
relations \cite{Berends:1987me}, adjusted to work for gluons in higher
dimensions. For efficiency, currents, which involve external gluons
only, are stored for re-use in evaluating other residues.

A number of consistency checks was carried out to verify the
correctness of the implementation. The gauge invariance of the results
and their independence of different choices for loop momenta and
dimensionalities $D$\/ and $D_s$ were tested. Coefficients themselves,
the pole structure of the amplitudes and, finally, the amplitudes
themselves have been compared to analytic results for various $N$\/
and different momentum and polarization configurations of the gluons.
Agreement within the limits of double-precision calculations has been
found with the numbers produced by Rocket for the fixed phase-space
points given in \cite{Giele:2008bc}.

In the following, studies are presented that have been conducted to
examine the accuracy and time dependence of the numerical calculation.
\smallskip

\begin{figure}[t!]
\centerline{
  \includegraphics[width=0.334\columnwidth]{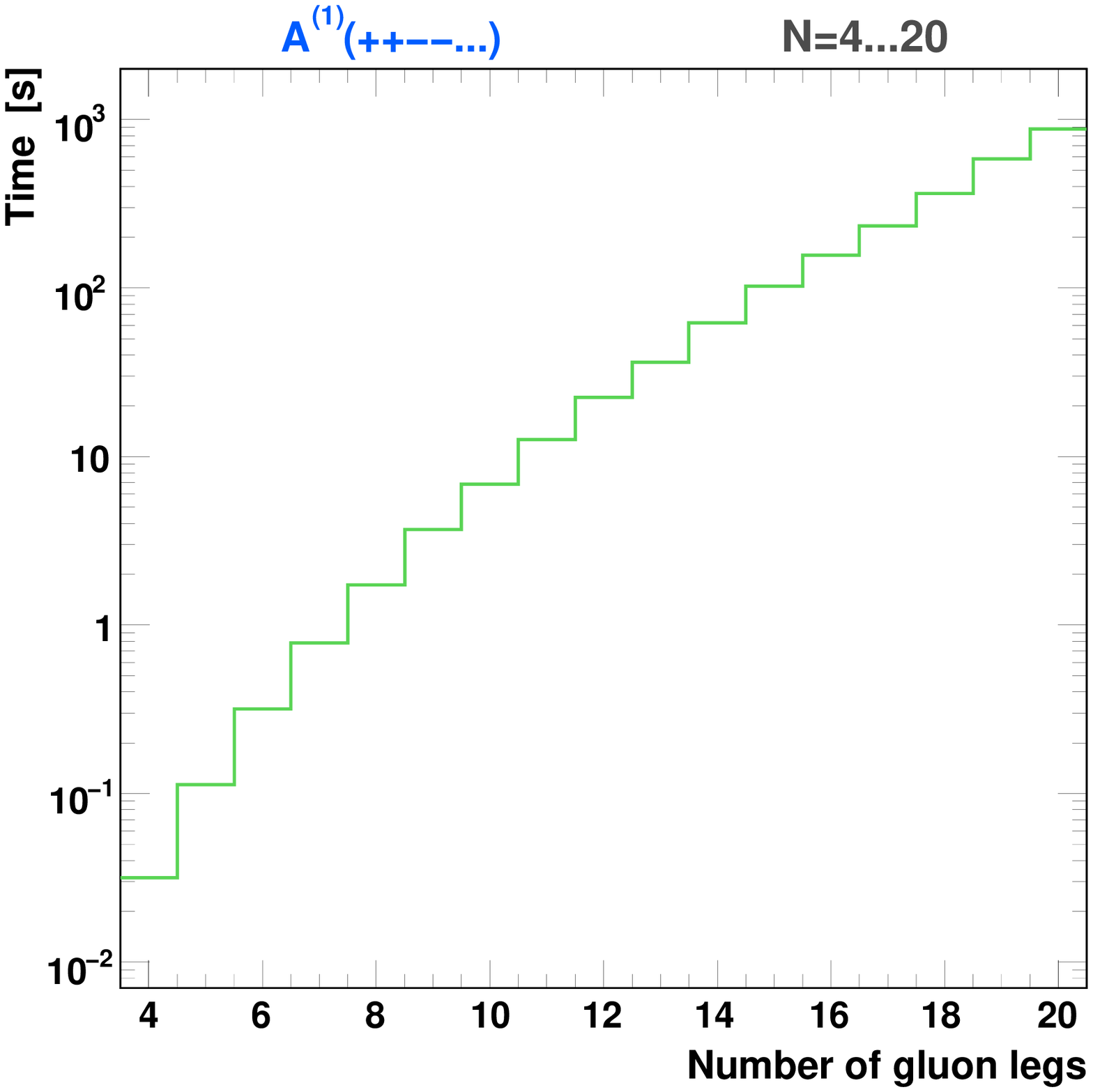}
  \includegraphics[width=0.334\columnwidth]{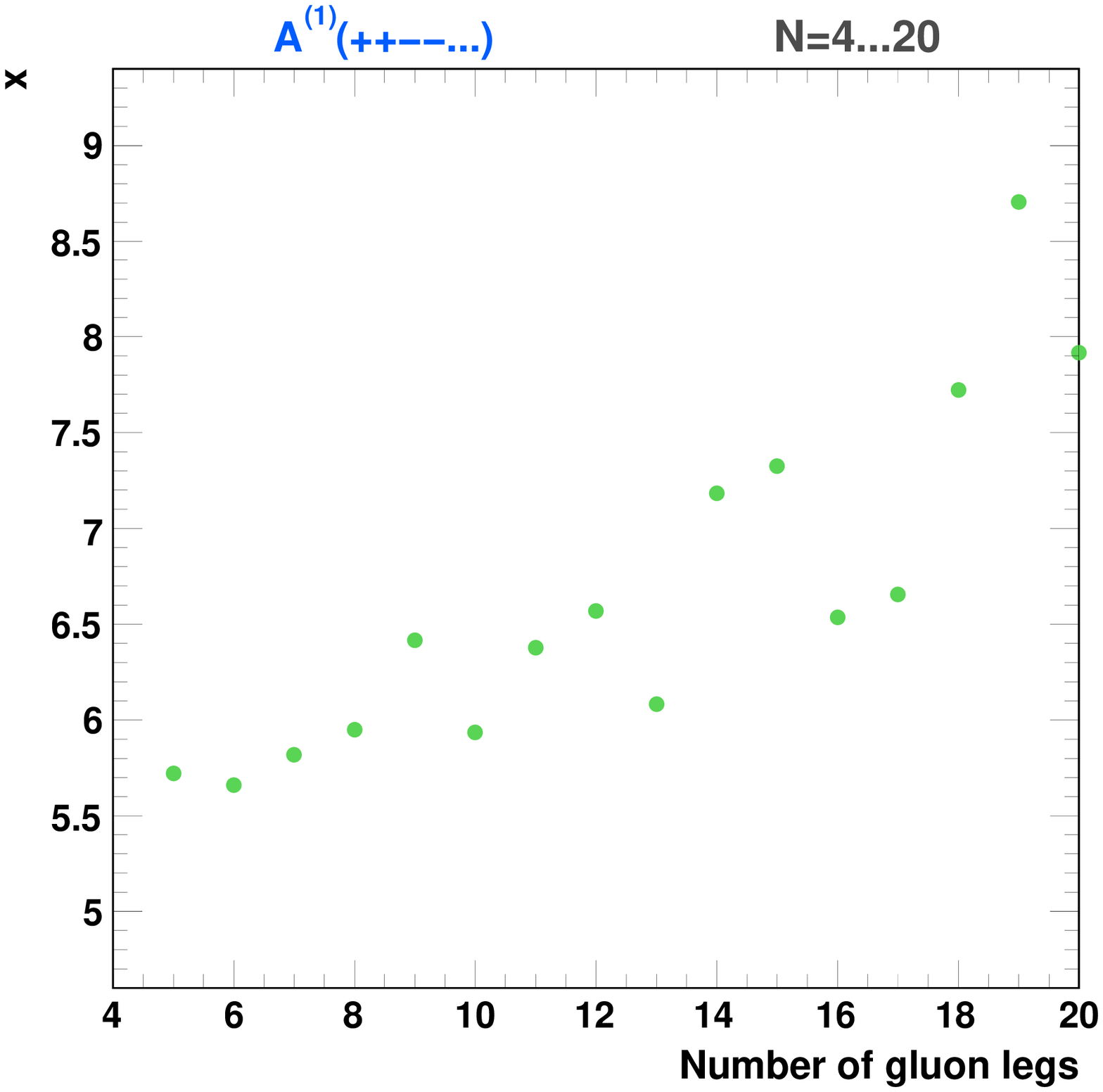}
  \includegraphics[width=0.334\columnwidth]{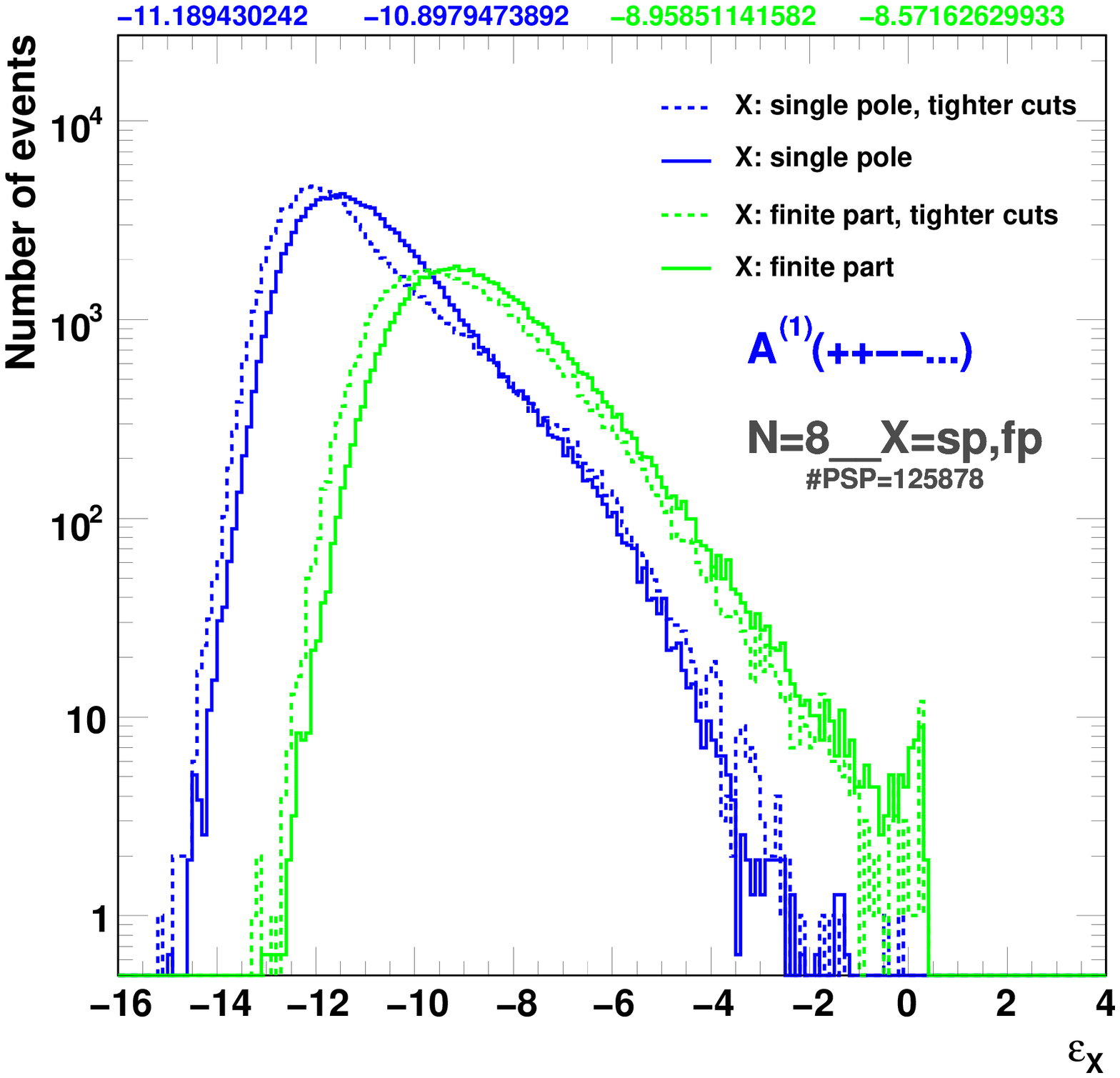}}
\caption{$N$ dependence of the computing time;
  $x_N$-exponents (center), see text. Times refer to using a 2.20 GHz
  Intel Core2 Duo processor.
  Tighter gluon cuts were used: $|\eta_i|<2$,
  $p_{\perp,i}\,s^{-0.5}>0.1$, $R_{ij}>0.7$, denoted as in
  \cite{Giele:2008bc}. The last plot shows the accuracy improvement.}
\label{Fig:speed}
\end{figure}

\noindent
{\bf Accuracy of the results.}\quad
\begin{figure}[t!]
\centerline{
  \includegraphics[width=0.34\columnwidth]{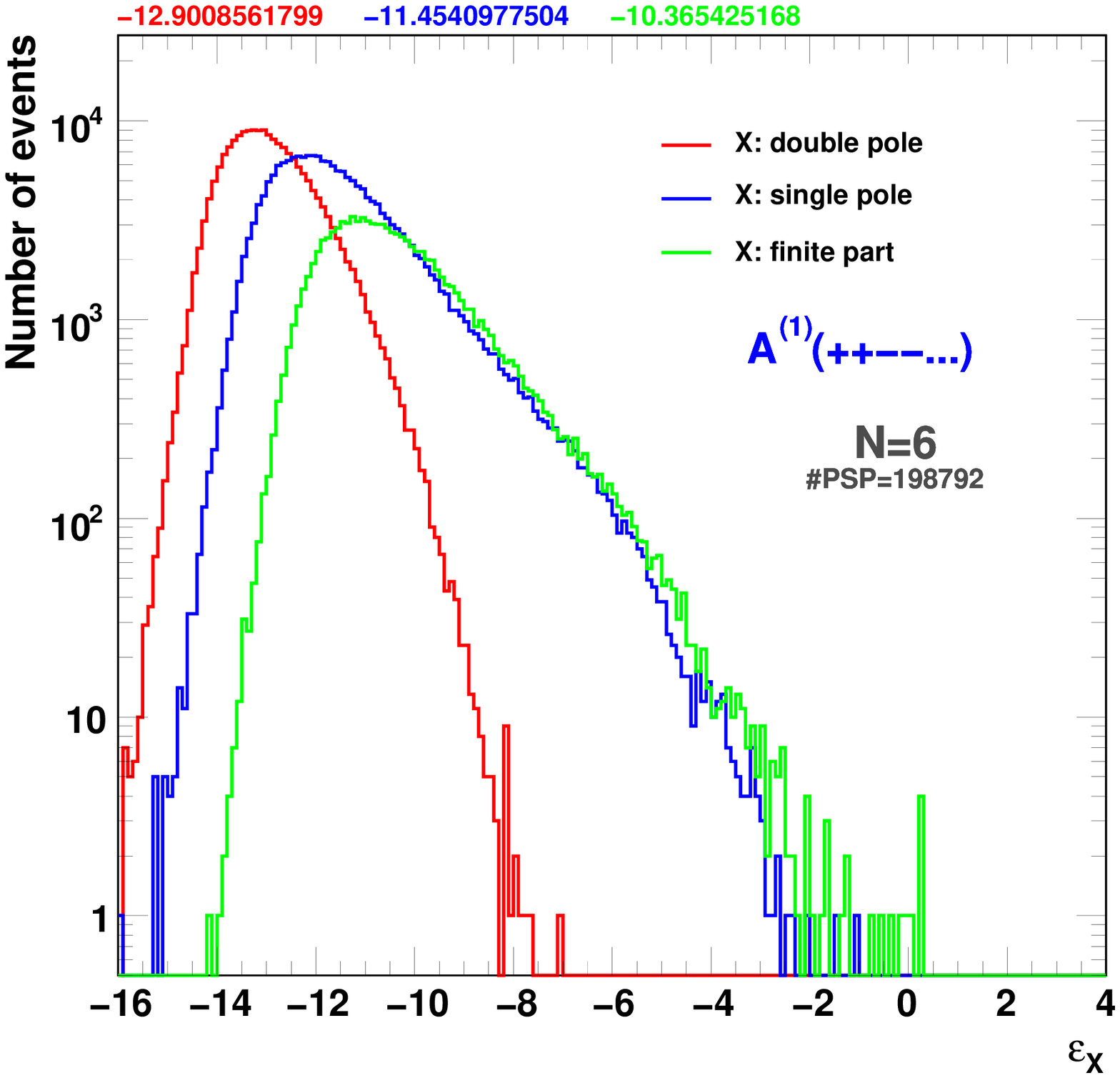}
  \includegraphics[width=0.34\columnwidth]{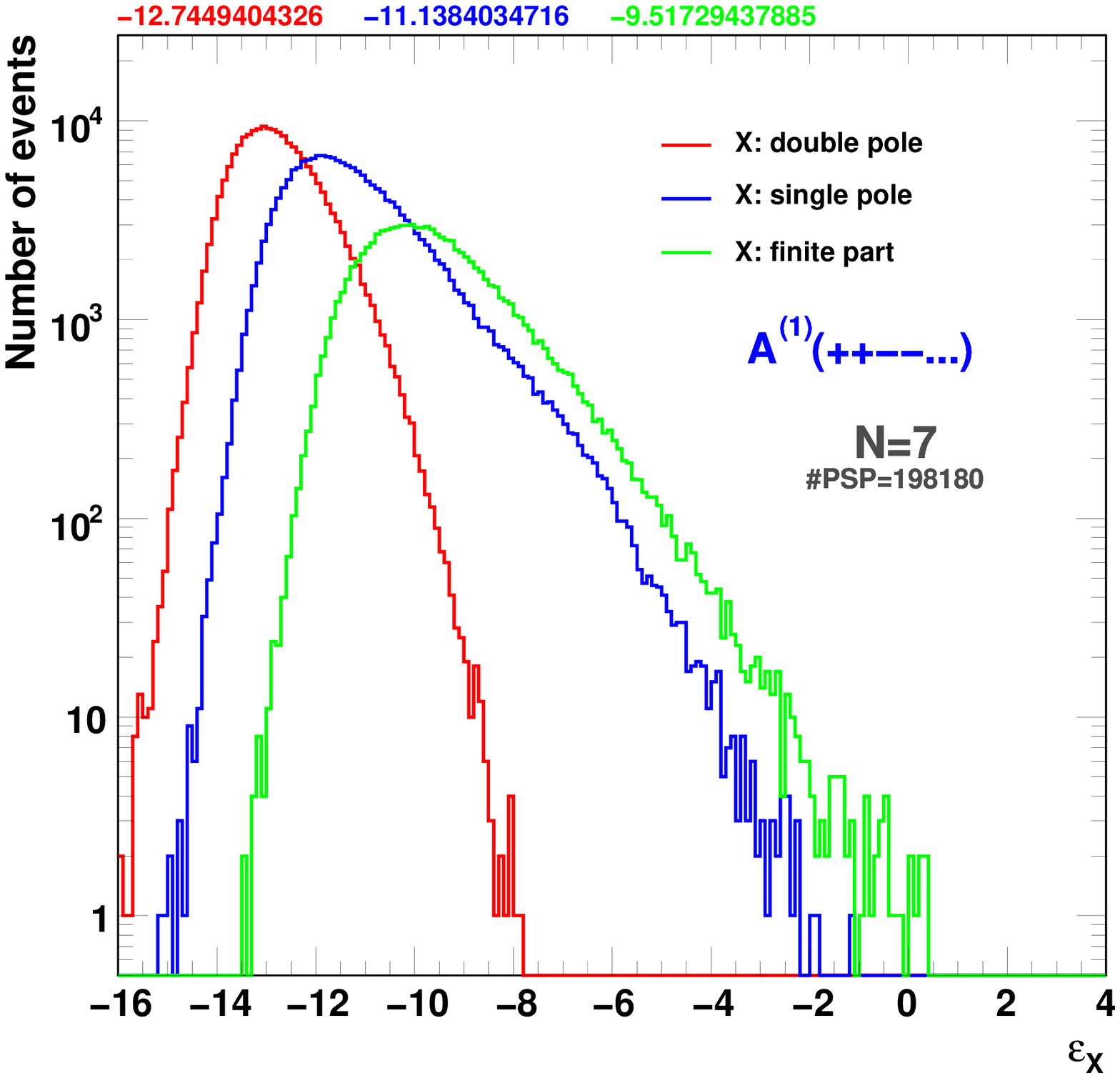}
  \includegraphics[width=0.34\columnwidth]{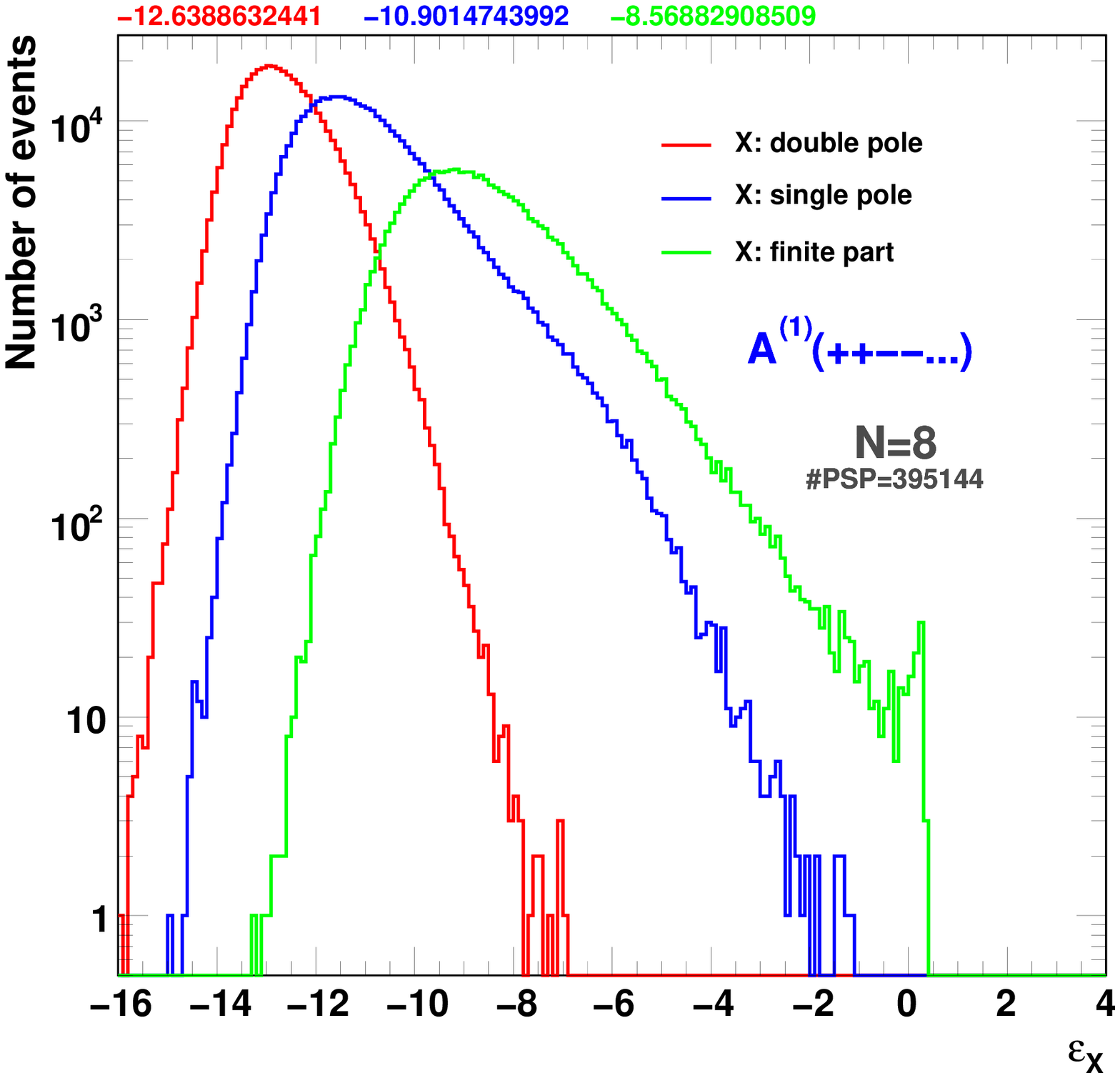}}
\centerline{
  \includegraphics[width=0.34\columnwidth]{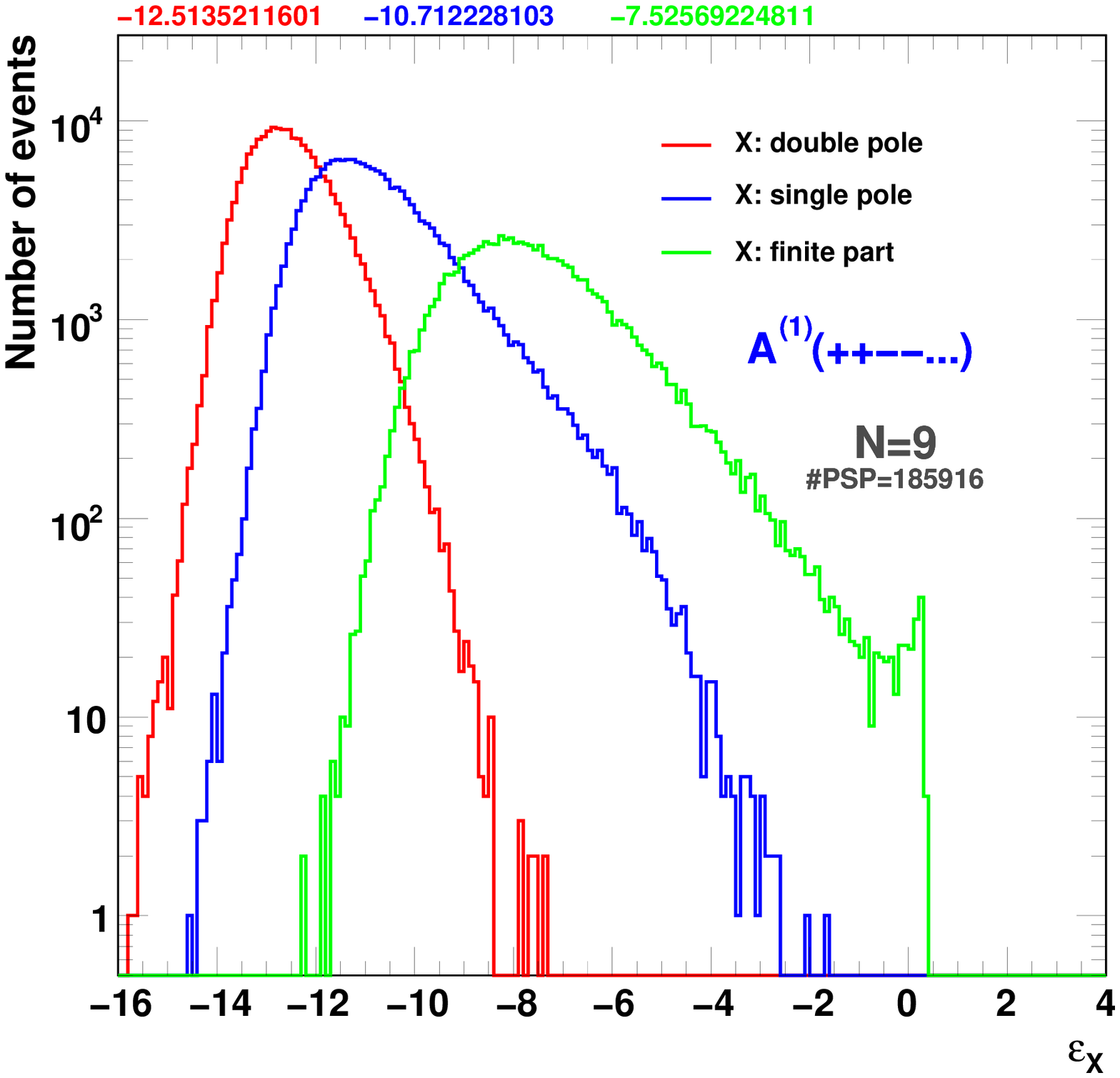}
  \includegraphics[width=0.34\columnwidth]{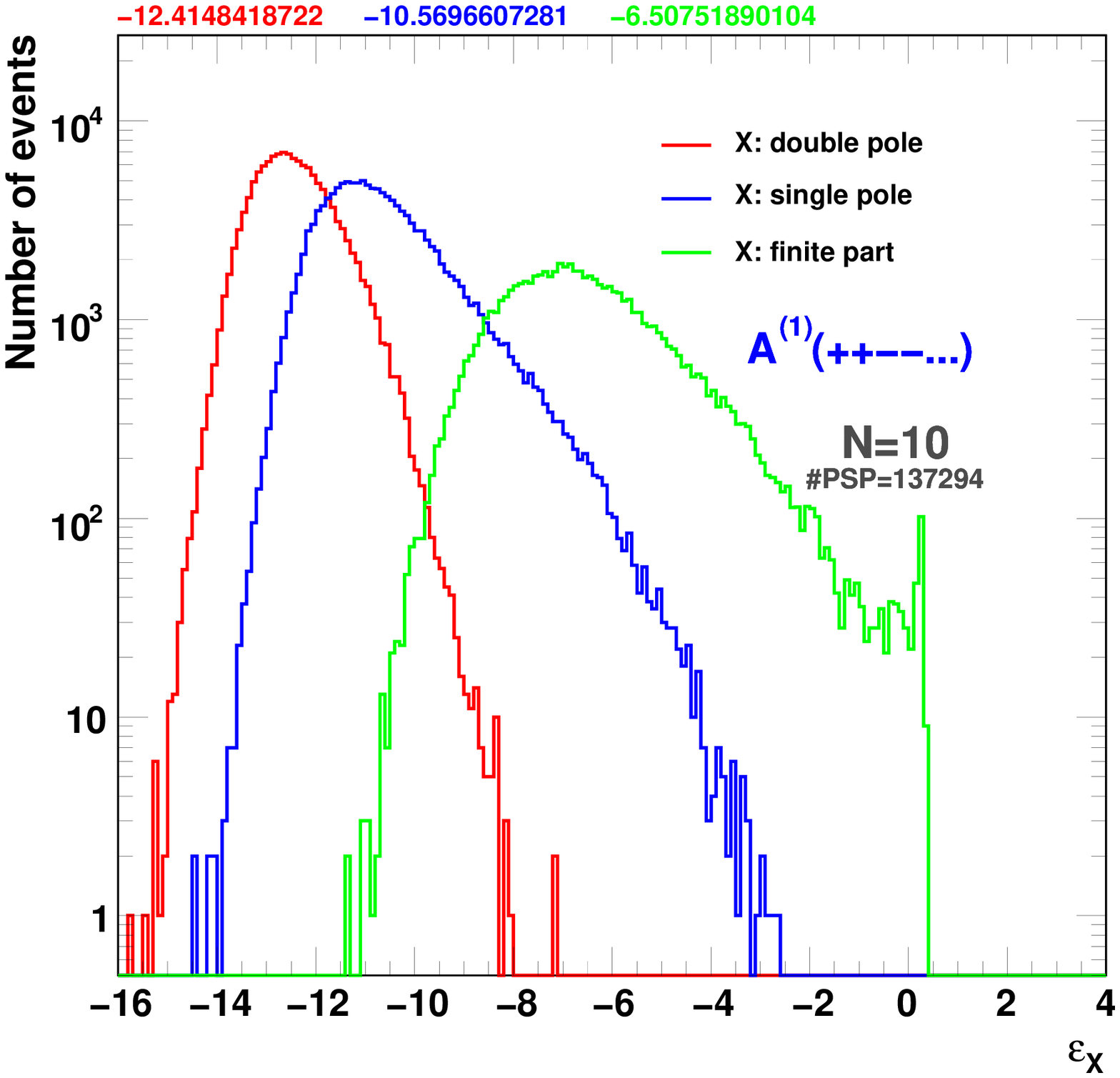}
  \includegraphics[width=0.34\columnwidth]{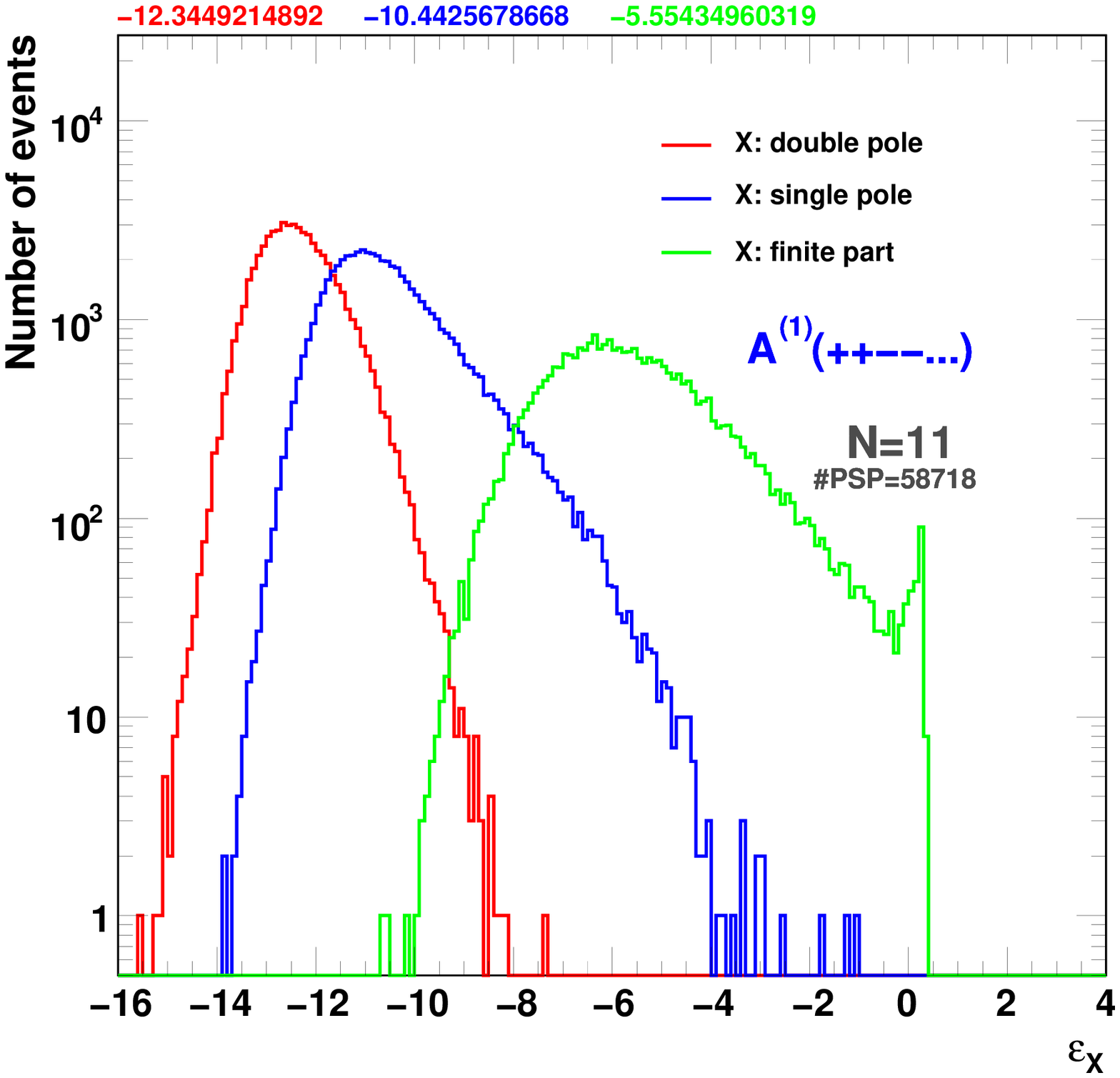}}
\centerline{
  \includegraphics[width=0.34\columnwidth]{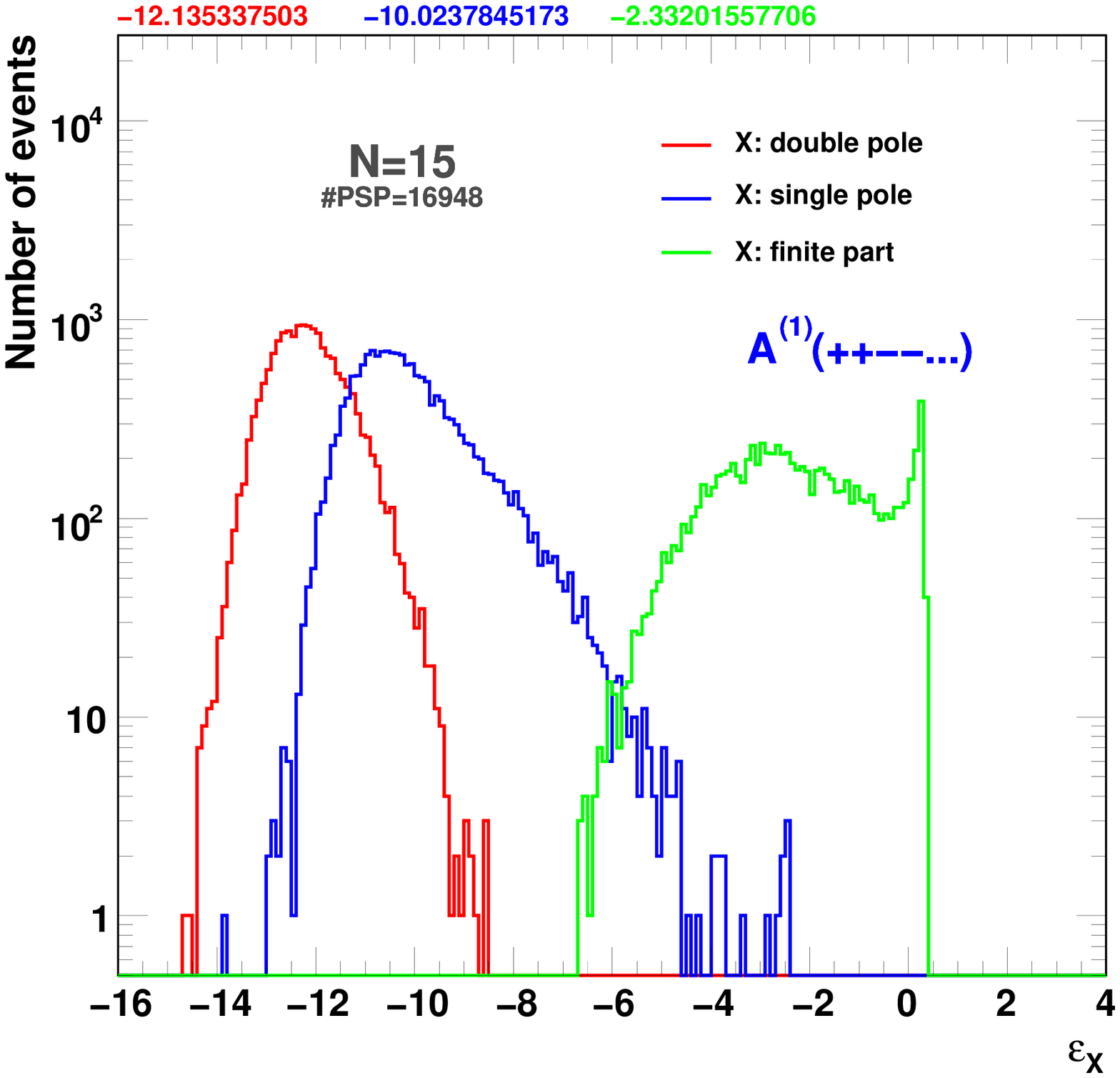}
  \includegraphics[width=0.34\columnwidth]{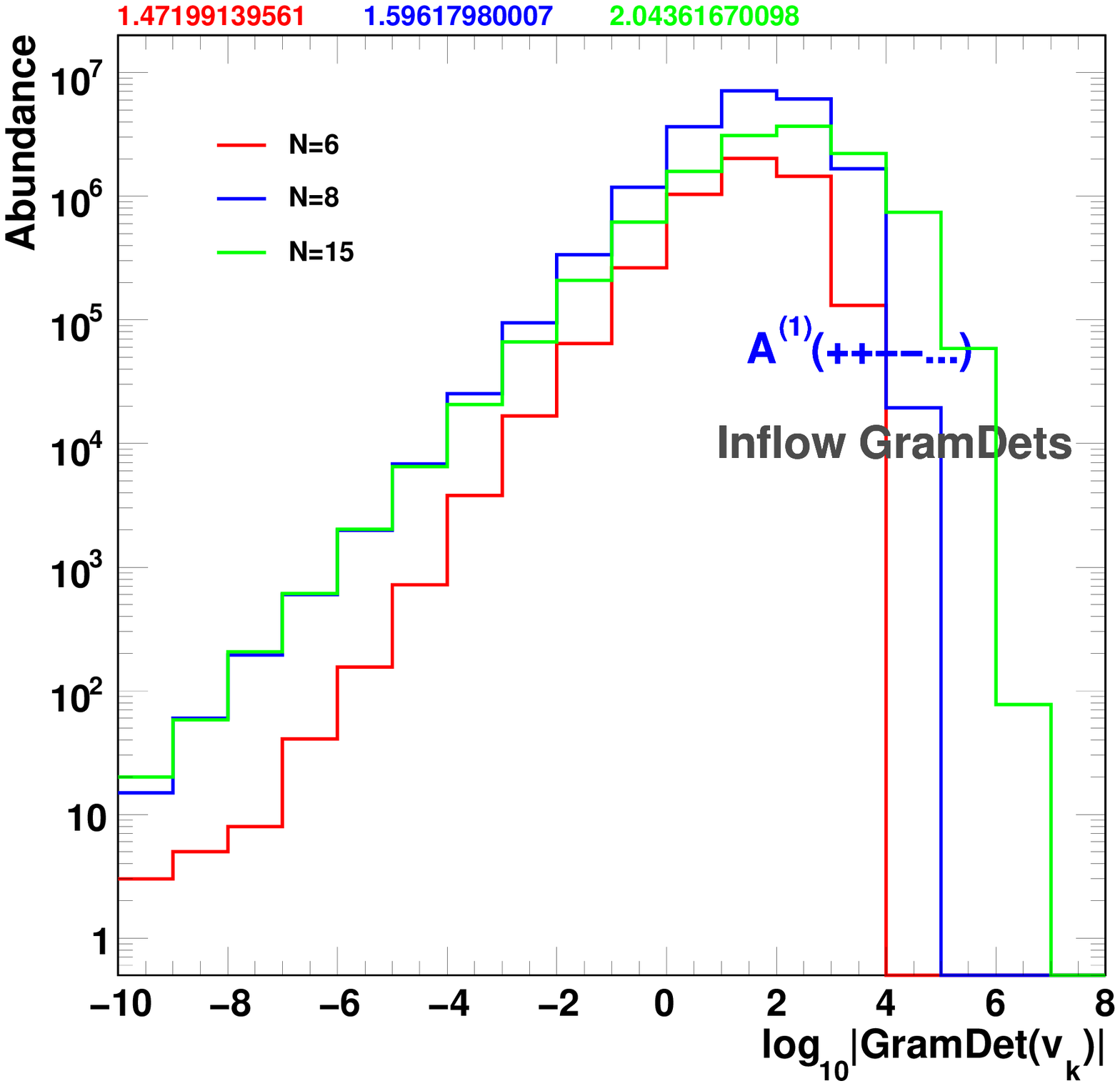}
  \includegraphics[width=0.34\columnwidth]{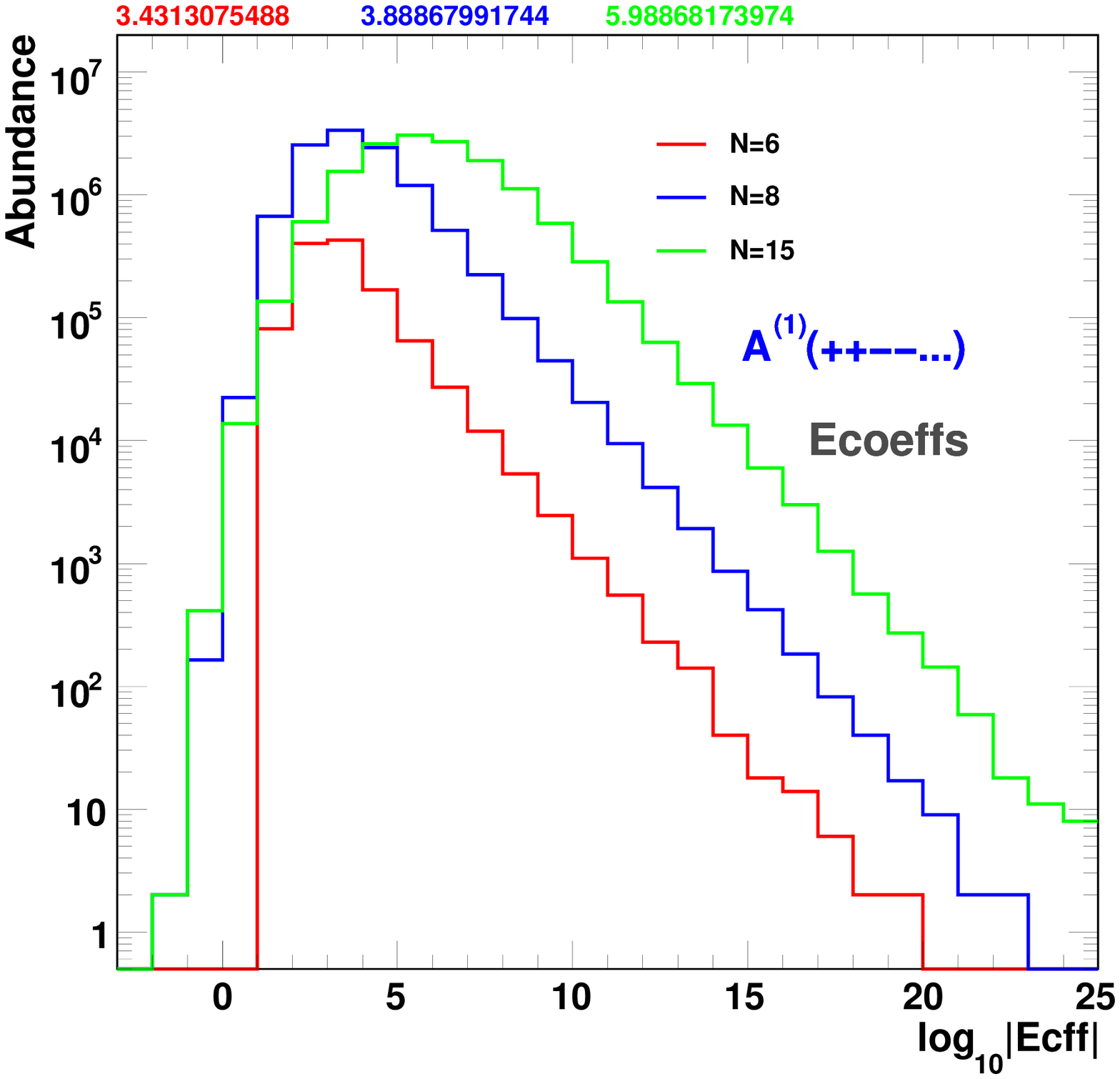}}
\caption{Double-, single-pole and finite-part accuracy (in double
  precision) of the $++--\ldots$ one-loop amplitudes for
  $N=6,\ldots,11,15$ gluons; see also text and right panel of
  Figure~\ref{Fig:speed}.
  Bottom row: center, double-logarithmic distributions of Gram
  determinants involving sets of external gluons and, right,
  $e^{(0)}_{i_1\cdots i_5}$ coefficients for the $N=6,8,15$ gluon
  setups.}
\label{Fig:accs+range}
\end{figure}
\begin{figure}[t!]
\centerline{
  \includegraphics[width=0.334\columnwidth]{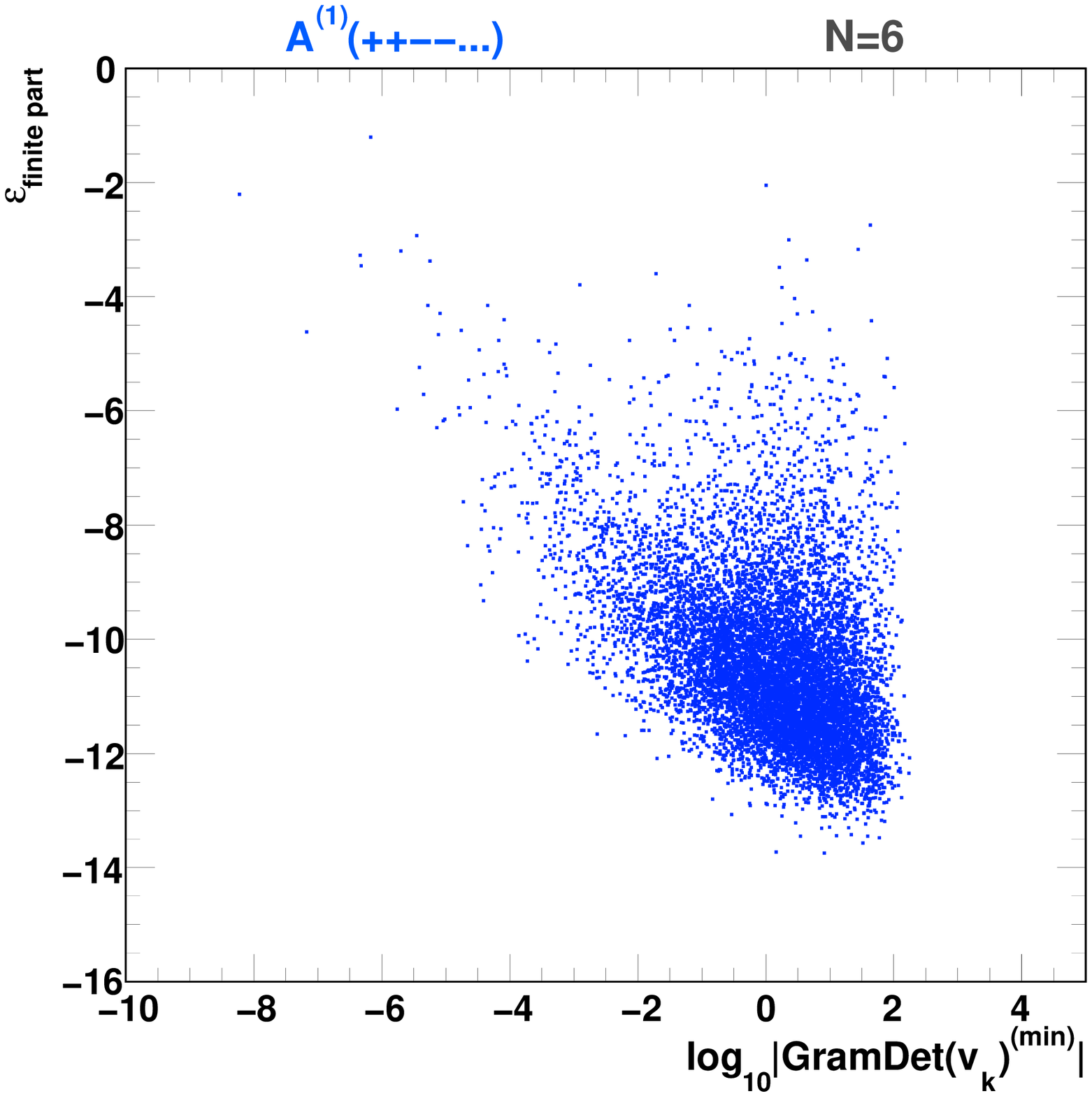}
  \includegraphics[width=0.334\columnwidth]{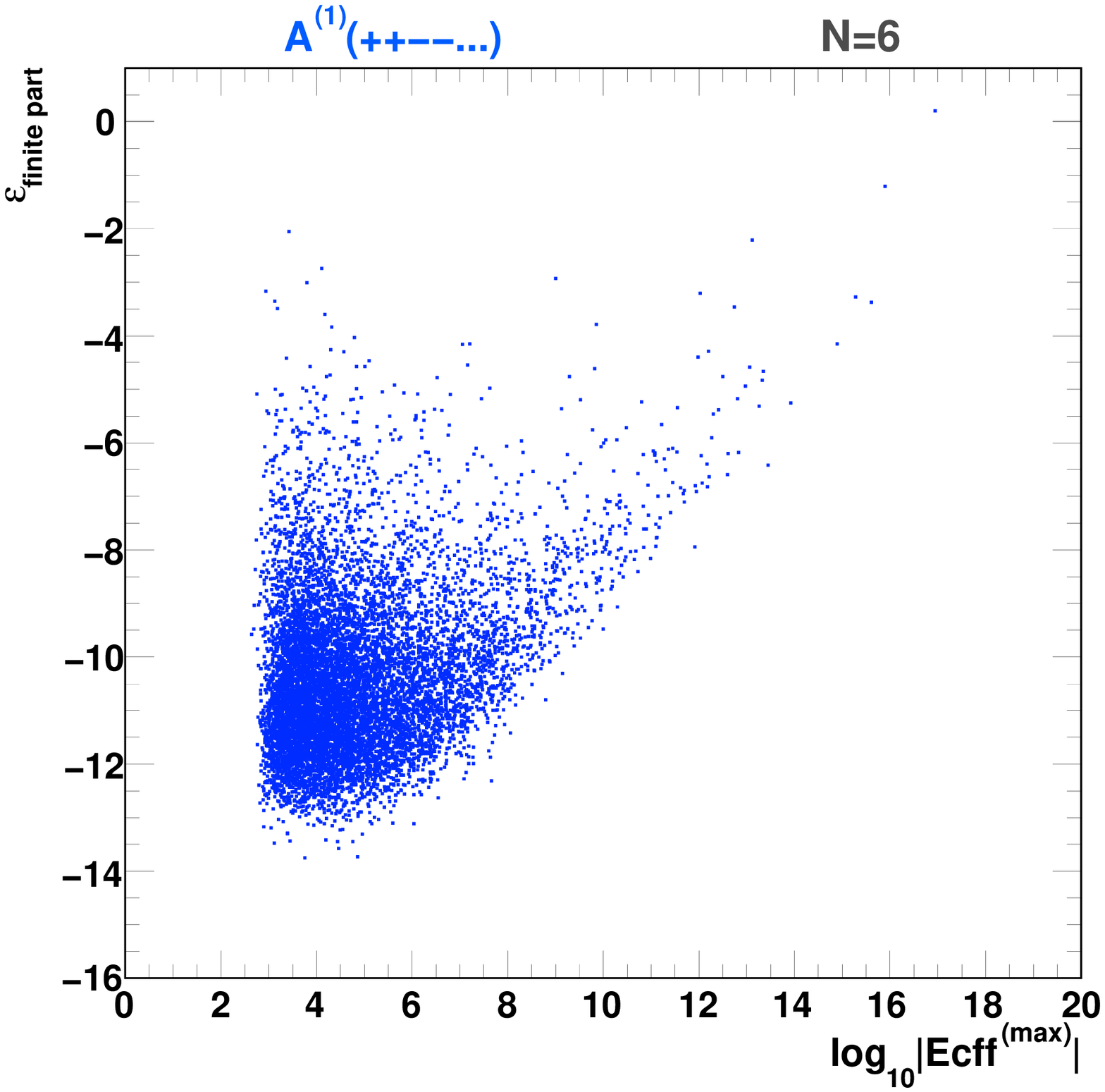}
  \includegraphics[width=0.334\columnwidth]{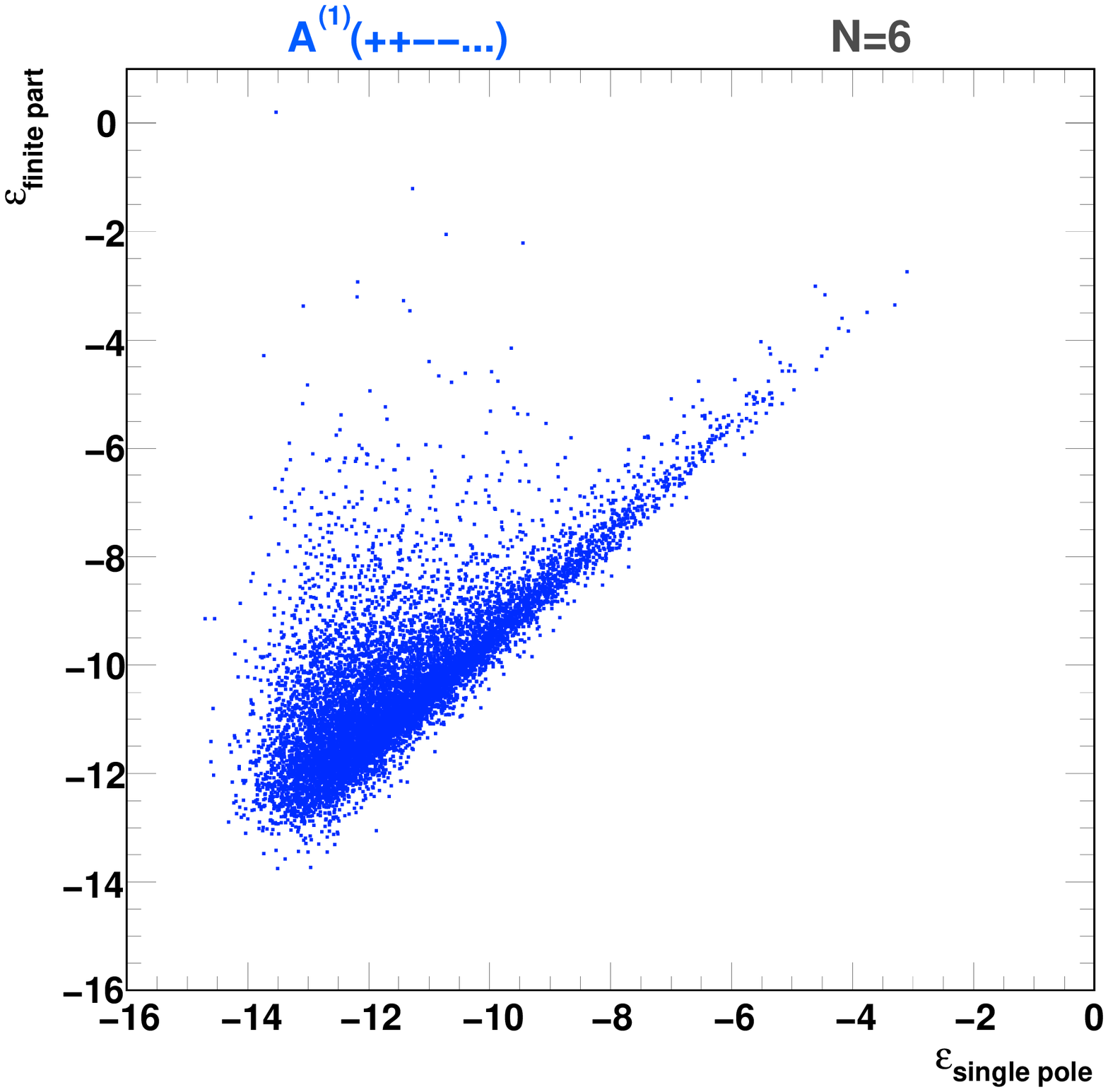}}
\centerline{
  \includegraphics[width=0.334\columnwidth]{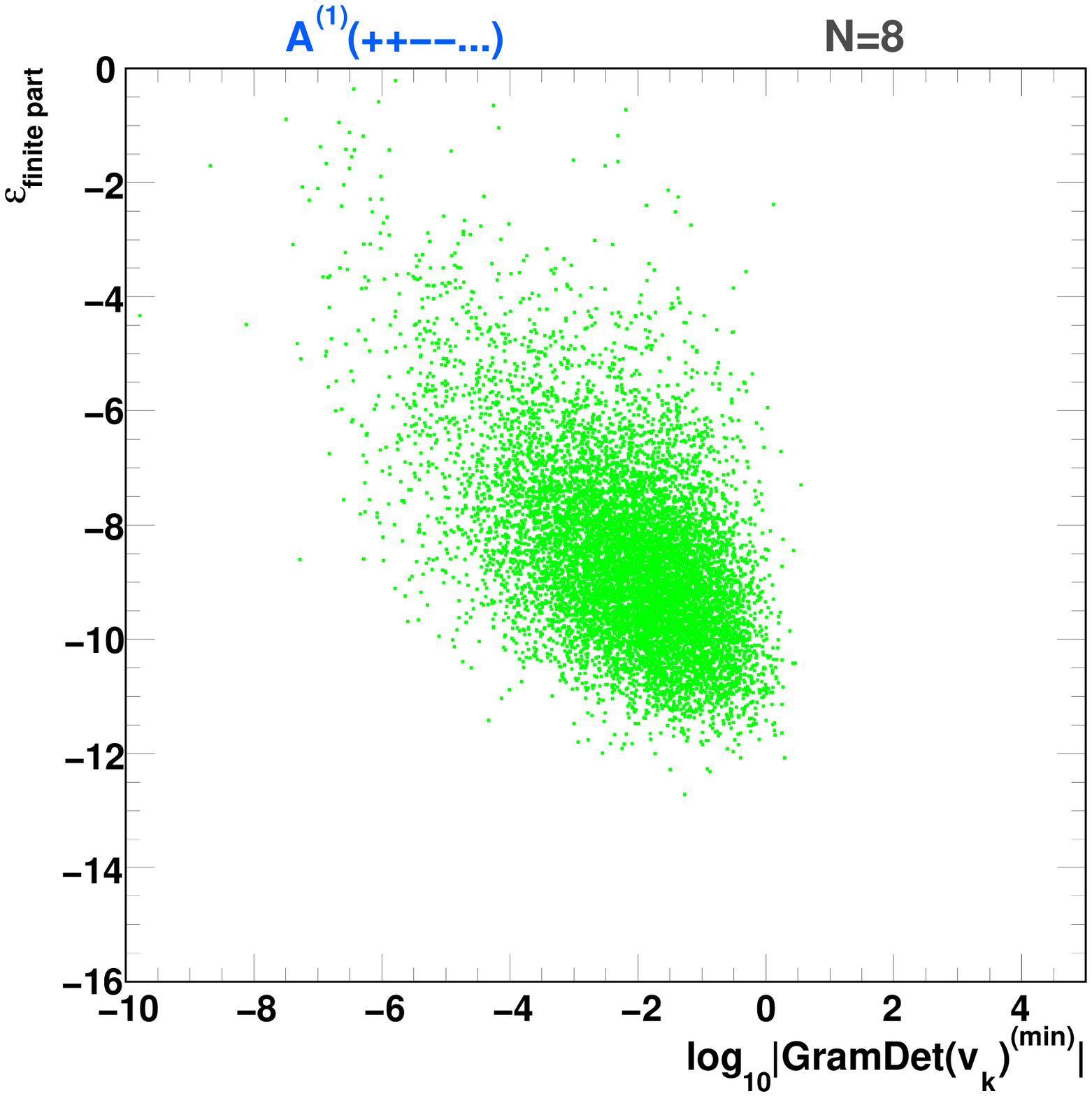}
  \includegraphics[width=0.334\columnwidth]{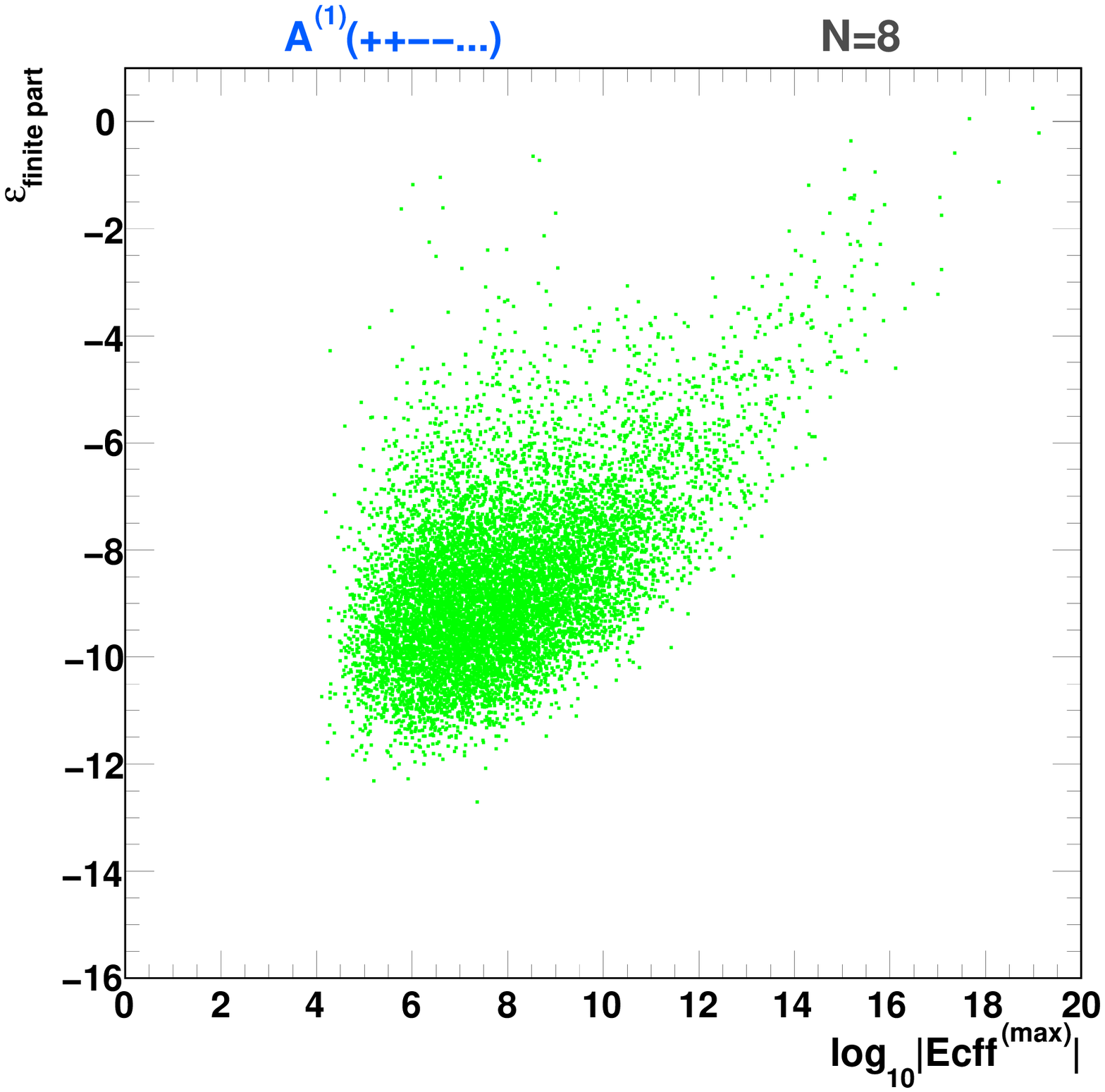}
  \includegraphics[width=0.334\columnwidth]{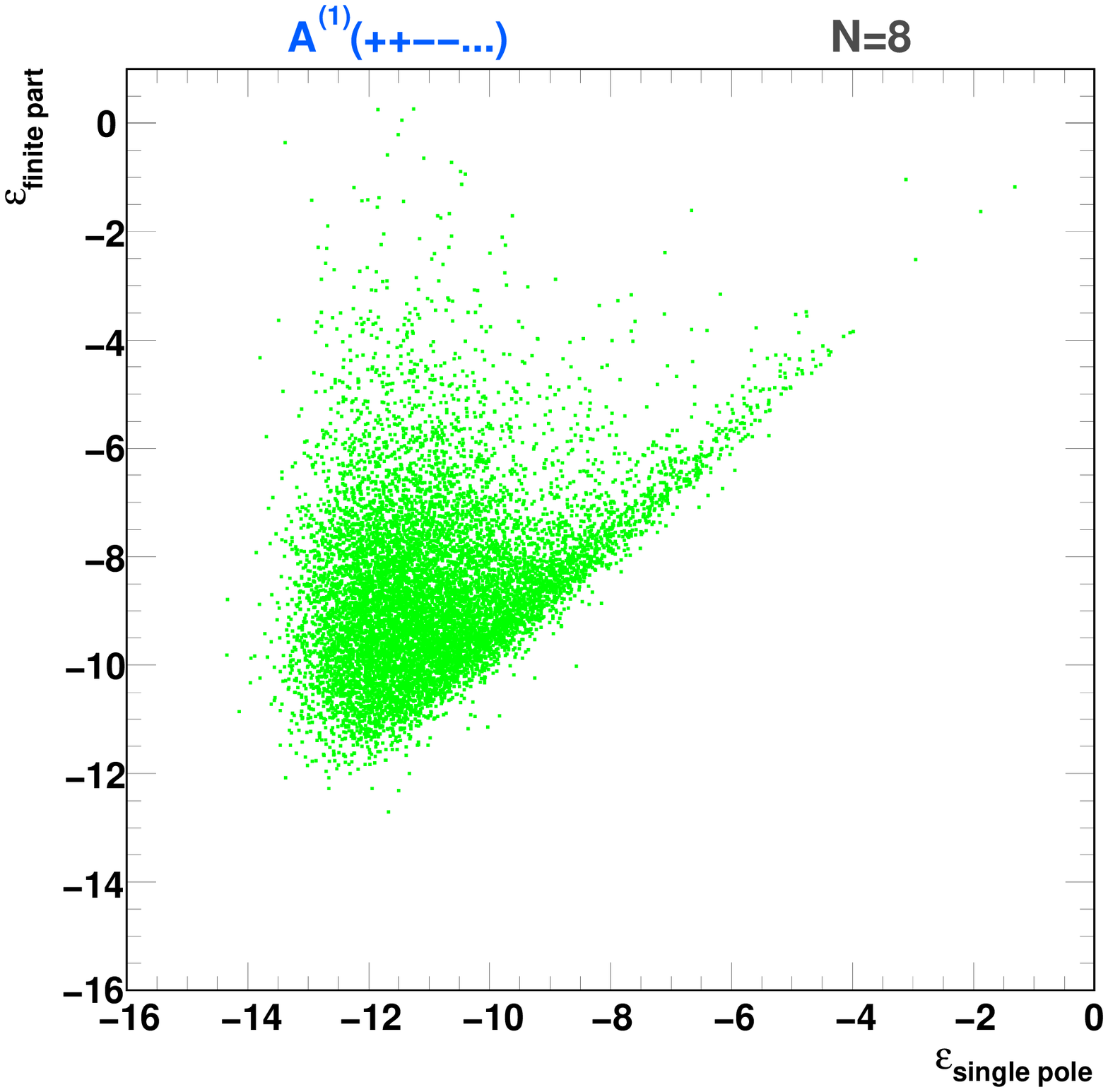}}
\centerline{
  \includegraphics[width=0.334\columnwidth]{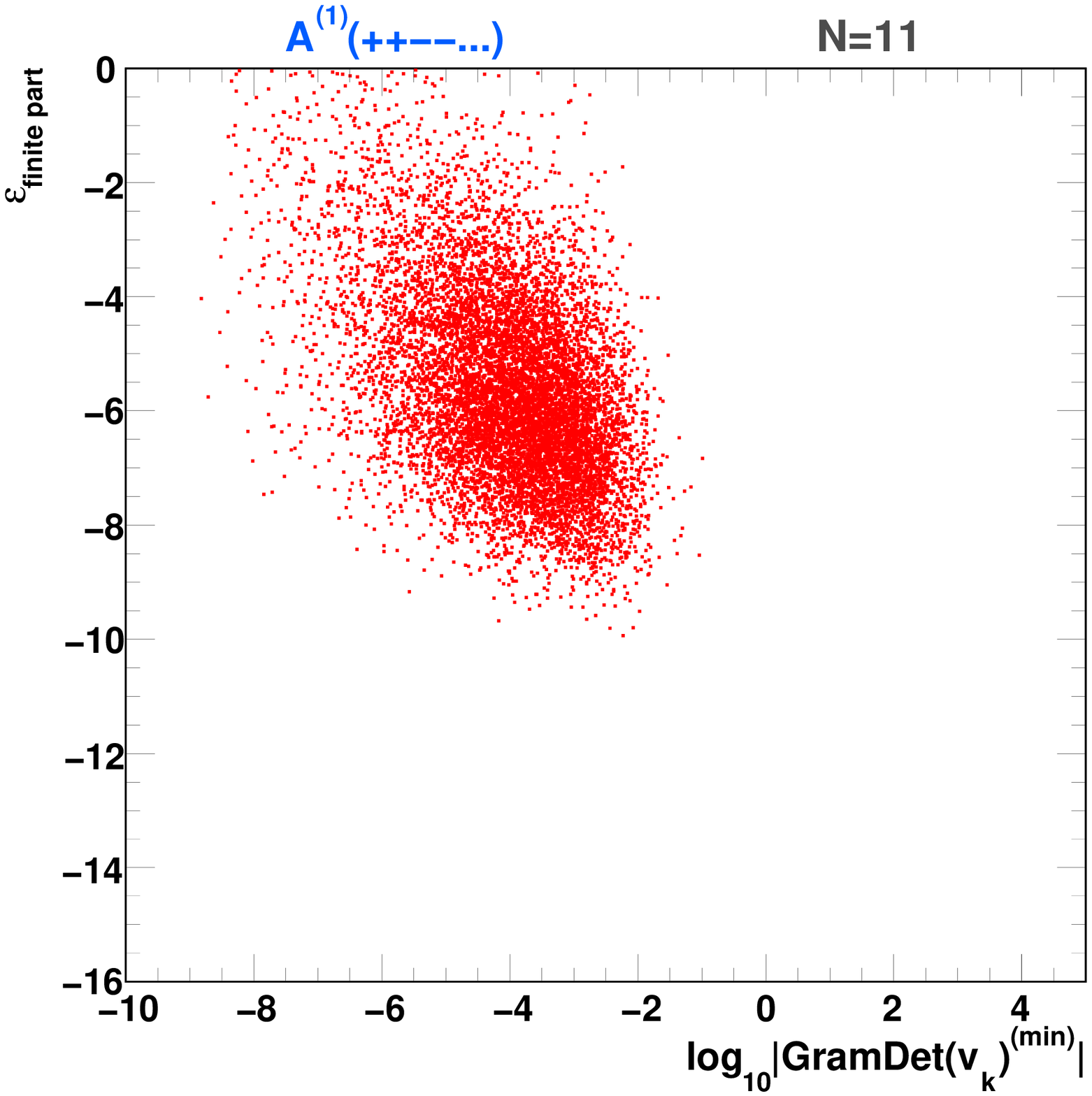}
  \includegraphics[width=0.334\columnwidth]{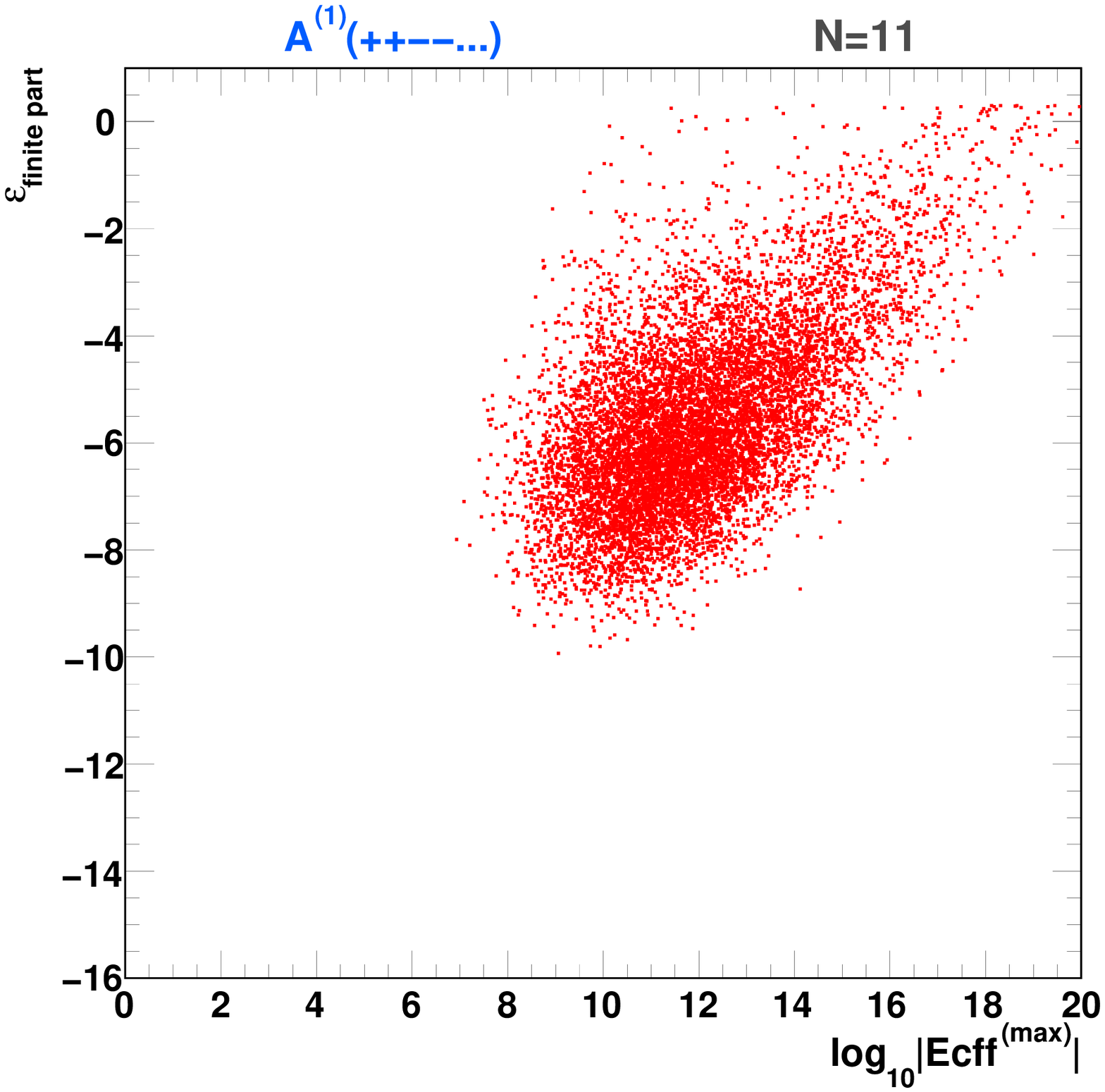}
  \includegraphics[width=0.334\columnwidth]{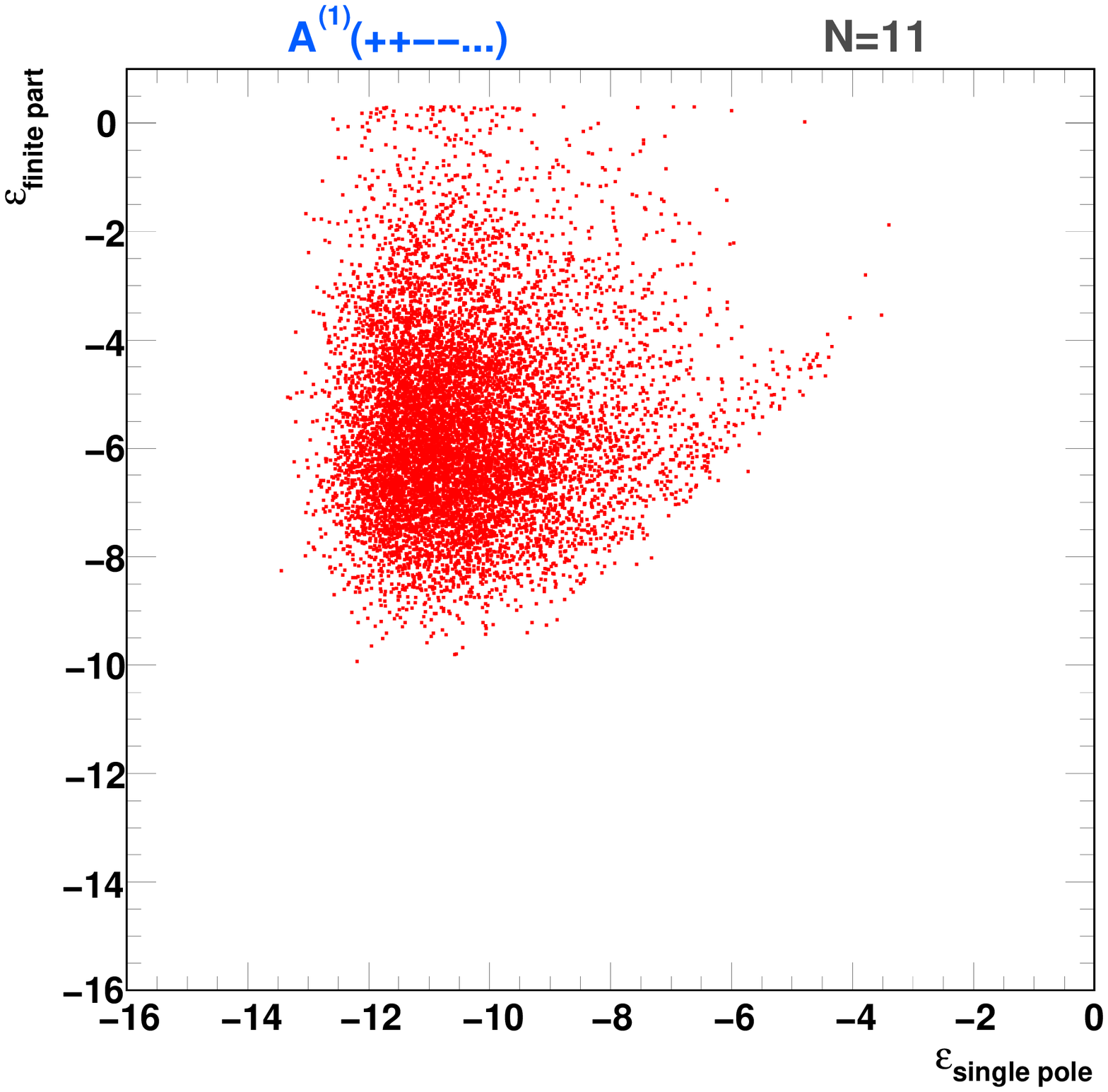}}
\caption{Finite-part accuracy versus minimal Gram determinant of
  external-gluon sets (left), maximal $e^{(0)}_{i_1\cdots i_5}$
  coefficient (center) and single-pole accuracy (all in double
  precision) for the $N=6,8,11$ gluon setups of above.}
\label{Fig:corrs}
\end{figure}
The quality of the numerical solutions can be estimated by analyzing
the logarithmic relative deviations, which are defined as
\begin{equation}
  \varepsilon_{\rm{dp,sp}}\;=\;\log_{10}
  \frac{|A^{[1](\rm{dp,sp})}_{N,\rm{C++}}-
    A^{[1](\rm{dp,sp})}_{N,\,\rm{anly}}|}
       {|A^{[1](\rm{dp,sp})}_{N,\,\rm{anly}}|}\,,\ \quad
  \varepsilon_{\rm{fp}}\;=\;\log_{10}
  \frac{2\,|A^{[1](\rm{fp})}_{N,\rm{C++}}[1]-
    A^{[1](\rm{fp})}_{N,\rm{C++}}[2]|}
       {|A^{[1](\rm{fp})}_{N,\rm{C++}}[1]|+
	 |A^{[1](\rm{fp})}_{N,\rm{C++}}[2]|}\,,
\end{equation}
where the analytically known pole structures of the one-loop
amplitudes are taken as reference for double (dp) and single (sp)
poles, while for finite parts (fp), two independent solutions are
compared with each other. Figure~\ref{Fig:accs+range} shows the
$\varepsilon$\/ distributions together with the number of generated
phase-space points for various $N$. The top row of numbers in the
plots displays the means of the distributions. All results have been
obtained for the same cuts on external gluons as reported in
\cite{Giele:2008bc}; for the effect of tighter cuts, see
Figure~\ref{Fig:speed} (right).

For $N=15$, the double-precision evaluation of the coefficients
clearly is not sufficient to yield reliable finite-part result. The
loss of precision as $N$\/ increases is correlated with the more
frequent appearance of small denominators and large numbers
characteristic for the calculation. The two rightmost bottom panels of
Figure~\ref{Fig:accs+range} present two examples by depicting the
range of magnitude taken by Gram determinants, used to evaluate the
$V_{i_1\cdots i_M}$ vectors of external gluons, and
$e^{(0)}_{i_1\cdots i_5}$ coefficients, which can be of
${\cal O}(10^{24})$ for $N=15$. Double precision will then be
insufficient to make cancellations as they may occur e.g.\ in
eq.~(\ref{eq:dbar}) manifest.
The scatter plots of Figure~\ref{Fig:corrs} visualize that (partial)
correlations exist between the relative accuracy of the finite part
and the smallest Gram determinant of external-gluon sets, the largest
$e^{(0)}_{i_1\cdots i_5}$ coefficient and the single-pole accuracy. In
all cases, the areas of scatters shift with increasing $N$\/ towards
worse accuracy and more extreme values of Gram determinants and
$5$-point coefficients. The calculation may still involve other small
denominators, such as the leftover $d_j$ in the subtraction terms of
e.g.\ eq.~(\ref{eq:dbar}). This leads to instabilities (even for small
coefficients) and uncorrelated areas in the plots are populated. Given
these correlations, it can be seen that one way of achieving higher
accuracy is to compute the coefficients in quadrupole precision. This
has been pointed out in Ref.~\cite{Giele:2008bc}.
\smallskip

\noindent
{\bf Efficiency of the calculation.}\quad
As estimated in Ref.~\cite{Bern:2008ef} (p.\ 31), the algorithm is
expected to have polynomial complexity, see also \cite{Giele:2008bc}.
The computing time $\tau_N$\/ to calculate an ordered $N$-gluon
one-loop amplitude should scale as $N^x+\ldots$, with the leading term
having $x=9$ dominating the behaviour for large $N$. The results for
$\tau_N$\/ with $N=4,\ldots,20$ are shown in Figure~\ref{Fig:speed}
together with the exponents
$x_N=\ln\frac{\tau_N}{\tau_{N-1}}/\ln\frac{N}{N-1}$.
The plots demonstrate that the C++ algorithm as implemented displays
the predicted scaling.

\section{Conclusions}
It has been shown that the generalized unitarity method of
Refs.~\cite{Ellis:2007br,Giele:2008ve} can be implemented in a stable
and fast C++ program. The one-loop $N$-gluon amplitude has been used
as a testing ground. The next step is to integrate the C++ code in an
existing leading-order generator. The resulting upgraded generator
will be able to generate both virtual and bremsstrahlung contributions
for arbitrarily complex Standard Model processes. The final step
towards a full next-to-leading order Monte Carlo is to add the
necessary phase-space integrations. As was shown in
Ref.~\cite{Ellis:2009zw} the virtual matrix elements calculated with
the generalized unitarity method can be used in next-to-leading order
Monte Carlo programs.
\vspace{-0.4mm}



\begin{footnotesize}

\end{footnotesize}


\end{document}